## Study of GEM-like detectors with resistive electrodes for RICH applications

A.G. Agócs, [1,2] A. Di Mauro,[3] A. Ben David,[4] B. Clark,[5] P. Martinengo,[3] E. Nappi,[6] V. Peskov[3, 7*]

[1]Eotvos University, Budapest, Hungary
[2]KFKI RMKI Research Institute for Particle and Nuclear Physics, Budapest, Hungary
[3]CERN, Geneva, Switzerland
[4]Tel Aviv University, Israel
[5]North Carolina State University, USA
[6]INFN Sezione di Bari, Bari, Italy
[7]Ecole Superior des Mines, St Etienne, France

### Abstract

We have developed prototypes of GEM-like detectors with resistive electrodes to be used as RICH photodetectors equipped with CsI photocathodes. The main advantages of these detectors are their intrinsic spark protection and possibility to operate at high gain (~$10^5$) in many gases including poorly quenched ones, allowing for the adoption of windowless configurations in which the radiator gas is also used in the chamber. Results of systematic studies of the resistive GEMs combined with CsI photocathodes are presented: its quantum efficiency, rate characteristics, long-term stability, etc. On the basis of the obtained results, we believe that the new detector will be a promising candidate for upgrading the ALICE RICH detector

### I. Introduction

Recent results from RHIC as well as numerous theoretical predictions indicate that a very high momentum particle identification (VHMPID) may be needed in the future upgrade of the ALICE experiments [1]. In connection to this the ALICE-HMPID collaboration is studying the possibility to build a new detector to identify charged particles with momentum above 5 GeV/c. Several Cherenkov detector designs were preliminary considered and simulated by the ALICE VHMPID group: a threshold- type as well as a RICH -type (see [2]). One of the complications in designing this new detector is the very limited space availability only about 1x1x1$m^3$ to locate the VHMPID in the present ALICE layout, so only rather compact and simple VHMPID designs can be considered. [2]. One of the promising photodetector elements in the VHMPID could be GEM-like detectors combined with CsI photocathodes. They have several advantages over other possible detectors: 1) they are compact and have planar geometry, 2) being coated with CsI layer they gain high quantum efficiency (QE) for UV, 3) can operate at higher gas gains and in badly quenched gases including inflammable gases, 4) can be used in the same gas as a Chereknkov radiator [3], thus no separating window between the radiator and the detector is needed, 5) can be made “hadron blind” [3], which is important when the contribution of the ionisation signal from charge particle should be suppressed et caetera.

* Corresponding author; Vladimir.peskov@cern.ch

For the last several years we were focused on developing more robust GEM-like detectors for RICH application. Our first successful prototype was an ”optimised” or “thick GEM” (TGEM) [4]. It allows to achieve gains almost ten times higher than with usual GEM thus offering a greater safety factor in operation. Preliminary study of TGEM for RICH application was performed in [5]. In this work we investigate a new promising candidate for RICH applications - resistive electrodes TGEMs, or RETGEM [6]. The main advantage of this novel detector is that it is fully spark-protected. The brief preliminary study indicates that another unique property of this detector is that if coated with a CsI photosensitive layer it gains high efficiency for UV, for example QE of 33% at 120 nm was already reported [6]. This opens the possibility to use it in RICH application. To confirm if RETGEM can be considered as a candidate for VHMPID detector in this work we focused on a systematic study of this detector: its QE, rate capability, short and long-term stability.

## Experimental set up

The experimental set up employed for studying the photosensitive RETGEM is shown schematically in Fig. 1.

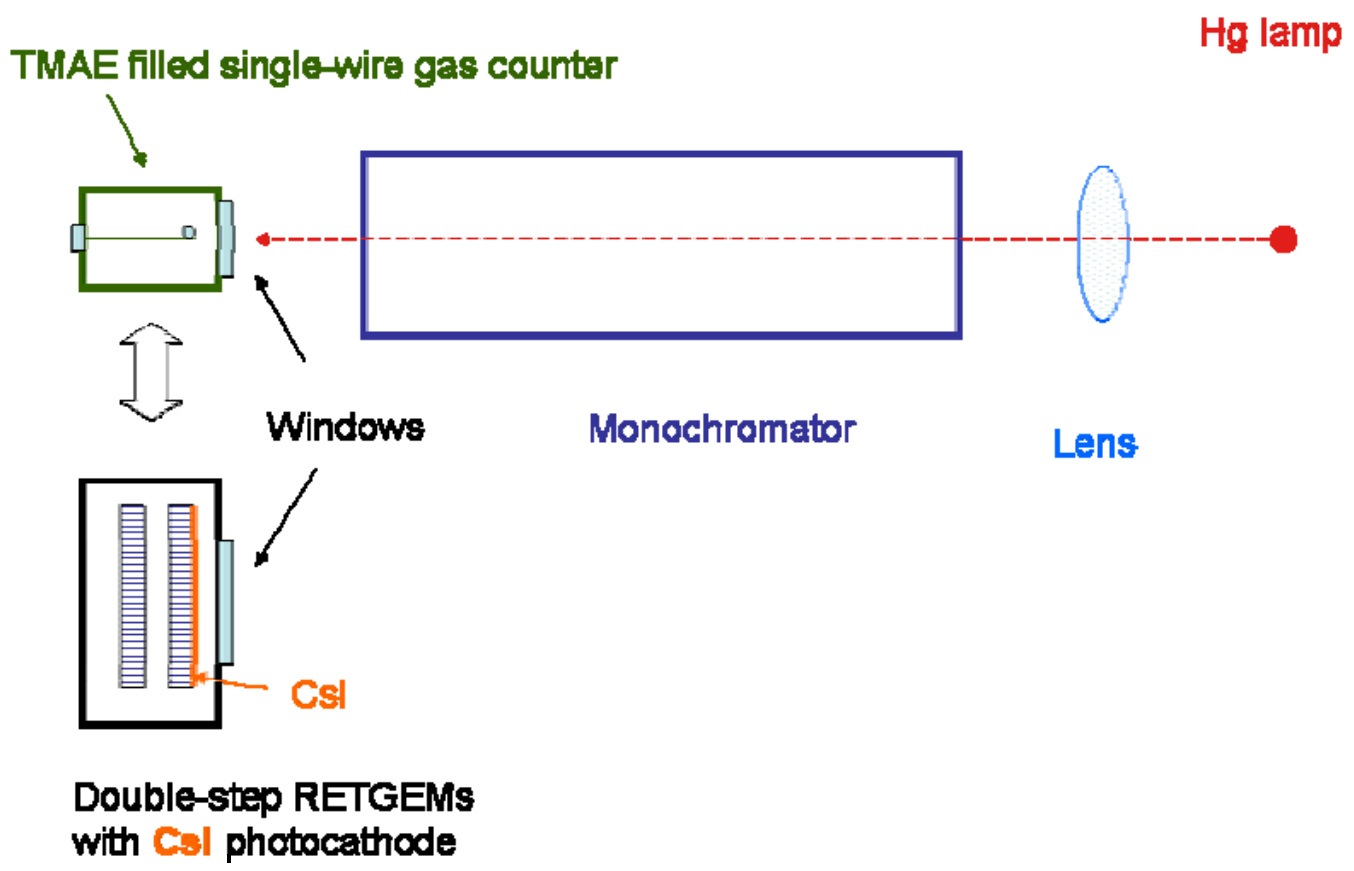

Fig.1. A schematic drawing of the experimental set up for the RETGEM study

It consists of a Hg lamp, a monochromator; a lens, focusing the light from the Hg lamp on the input slit of the monochromator and two gaseous detectors which can be placed close to the exit slit of the monochromator: a single- wire counter or a gas chamber with double RETGEM installed inside. The RETGEMs used in this work were manufactured from resistive Kapton 100XC10E5 (see [6] for more details) and had the following geometrical characteristics: thickness 1 mm; hole diameters of 0.5 mm, pitch 0.8 mm. The top electrode of the double RETGEM was coated by a vacuum evaporation technique with 0.35 μm thick CsI. The single-wire counter was filled with Ar+10%$CO_2$+TMAE gas mixture at a total pressure of p=1 atm and the gas chamber containing RETGEM was flushed with one of the following gases: Ne, Ar or $Ar+10\%CO_2$ also at p=1 atm. In the case of the single- wire counter the UV light from the monochromator caused the photoionization of TMAE vapours (the depth of the active part of this detector was 4 cm, so almost full absorption of the UV light occurs inside its sensitive volume) and created photoelectrons triggered Townsend avalanches near the anode wire. The double RETGEM worked on the principle of the surface photoeffect: the UV light liberated photoelectrons from the CsI layer and these electrons triggered avalanches in RETGEM holes. The avalanche signals from both detectors were recorded by a charge -sensitive amplifier (Ortec or CAMBERRA) and if necessary were further amplified by a research amplifier. The QE of the RETGEM $Q_{CsI}$ was calculated from the following formula:

$$Q_{CsI}=Q_{TMAE}N_{CsI}/N_{TMAE}, \quad (1),$$

where $Q_{TMEA}$ is TMAE QE, $N_{TMAE}$, and $N_{CsI}$- are counting rates from the single- wire counter and from the RETGEM respectively.

In some stability measurements the RETGEM was irradiated directly by Hg lamp (without the monochromator); in this case the intensity of the lamp was attenuated with the help of filters.

In some control measurements, for example counting plateau measurements, in addition to the UV light 6 keV X-ray photons from $^{55}Fe$ were used.

## Results

For the correct evaluation of the $Q_{CsI}$ from formula (1) it is important, of course, that both detectors have a counting plateau indicating that full photoelecton collection is achieved in both detectors. This is the reason why our first measurements were dedicated to estimate the rate vs. voltages characteristics: $N_{TMAE}=N_{CsI}$ ($V_{CsI}$) and $N_{CsI}=N_{CsI}$ ($V_{RETGEM}$), where $V_{CsI}$ and $V_{RETGEM}$ are voltages applied to the single-wire detector and the RETGEM respectively. Some results are shown in Fig. 2a and b. It can be seen that in the voltage interval of $V_{CsI}$= 1880-1940V (Fig. 2a) and $V_{RETGEM}$=665-680V (Fig. 2b) the same type of plateau was observed, therefore for the $Q_{CsI}$ evaluation we assumed the counting rates values at $V_{CsI}$= 1940V and $V_{RETGEM}$= 675V.

In Fig. 3 are shown the spectra of the Hg lamp measured at these voltages with the single-wire counter and with the RETGEM. One can see a pick at 185 nm corresponding to the emission line of Hg. The ratio of the counting rates at the pick value was ~2 which gave the $Q_{CsI}$=14.5% for this particular case. One should note that due to the holes the open area of our RETGEMs was ~40%, thus the expected QE of the CsI coated surface without holes could be as high as 36%, indicating that the quality of the CsI photocathode evaporated on the top of the Kapton substrate was very good.

Fig. 4 shows typical results of the stability measurements obtained with the RETGEM, irradiated by the focused light beam (~6 $mm^2$) from the monochromator. As expected, due to the high resistivity of the Kapton the charging up effect was observed with time indicating that the $Q_{CsI}$, depending on conditions, could be in the interval of 12-14.5%. The simulations performed in [2] indicate that the $Q_{CsI} \geq 12\%$ should be sufficient for the mirror–based design of the VHPID detector; hence the already achieved efficiency allows to consider the RETGEM as a promising photodetector for low rate RICH detectors (as it should be the case of VHMPID).

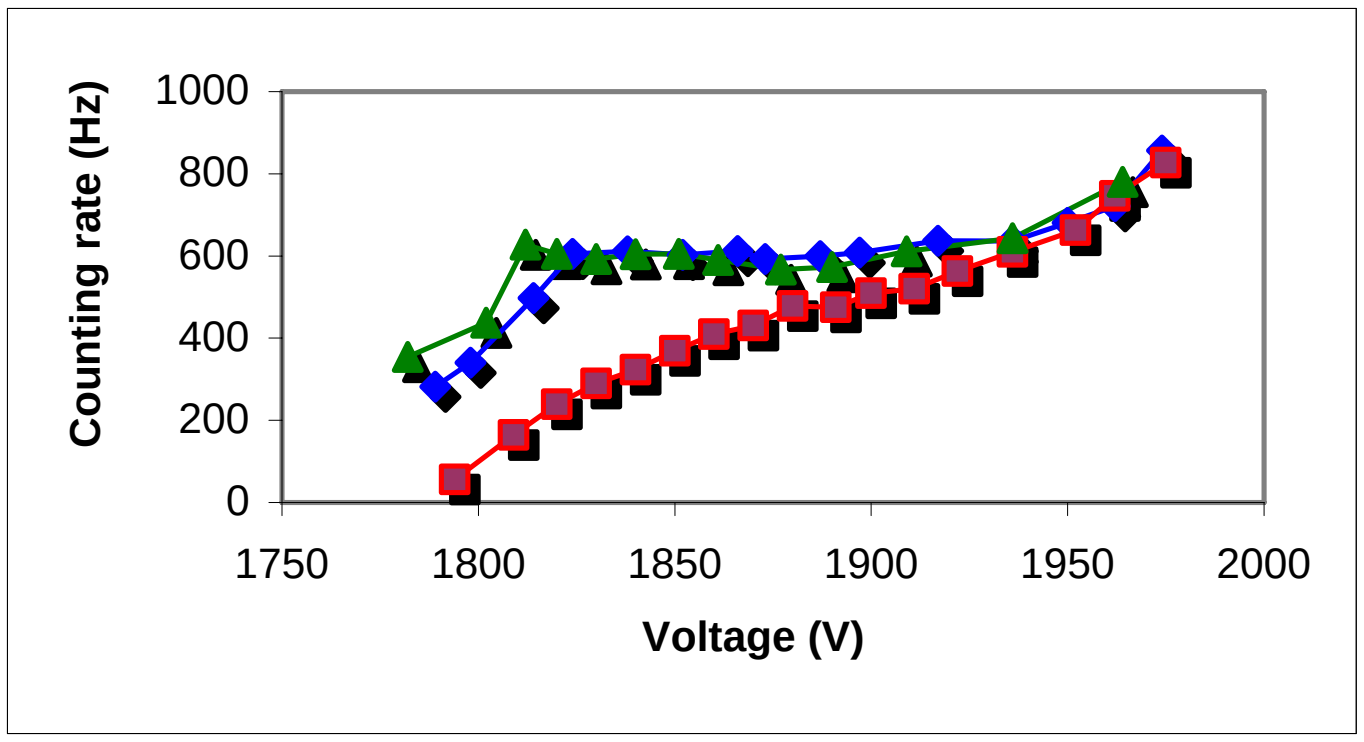

Fig. 2a. $N_{TMAE}$ vs. $V_{TMAE}$ measured in the case of the single-wire counter: 1- the UV light from the Hg lamp, 2,3-55Fe. Gas mixture Ar+10%CO2+TMAE=1 atm.

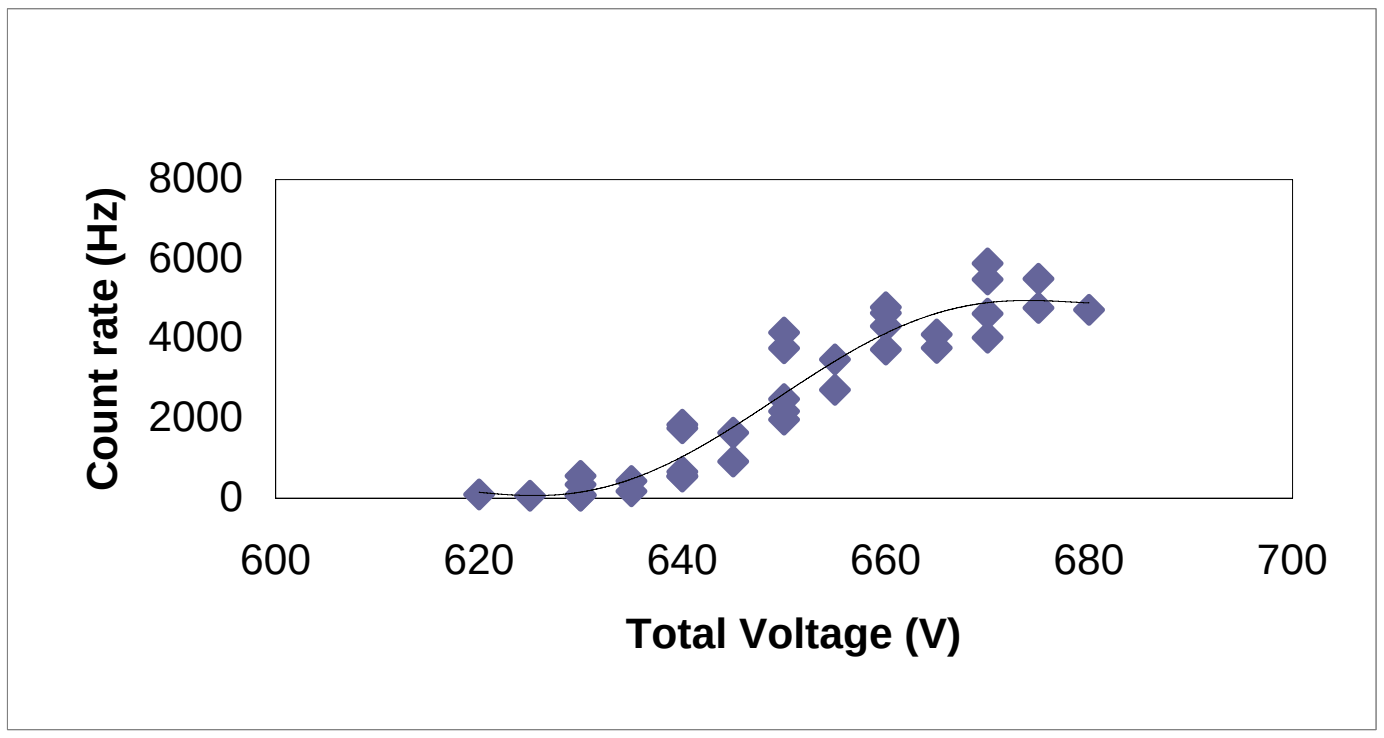

Fig. 2b. $N_{CsI}$ vs. $V_{RETGEM}$ for double RETGEM operating in Ne at p=1 atm. Similar results were obtained in the case of Ar and Ar+10%$CO_2$, however at considerably higher voltages

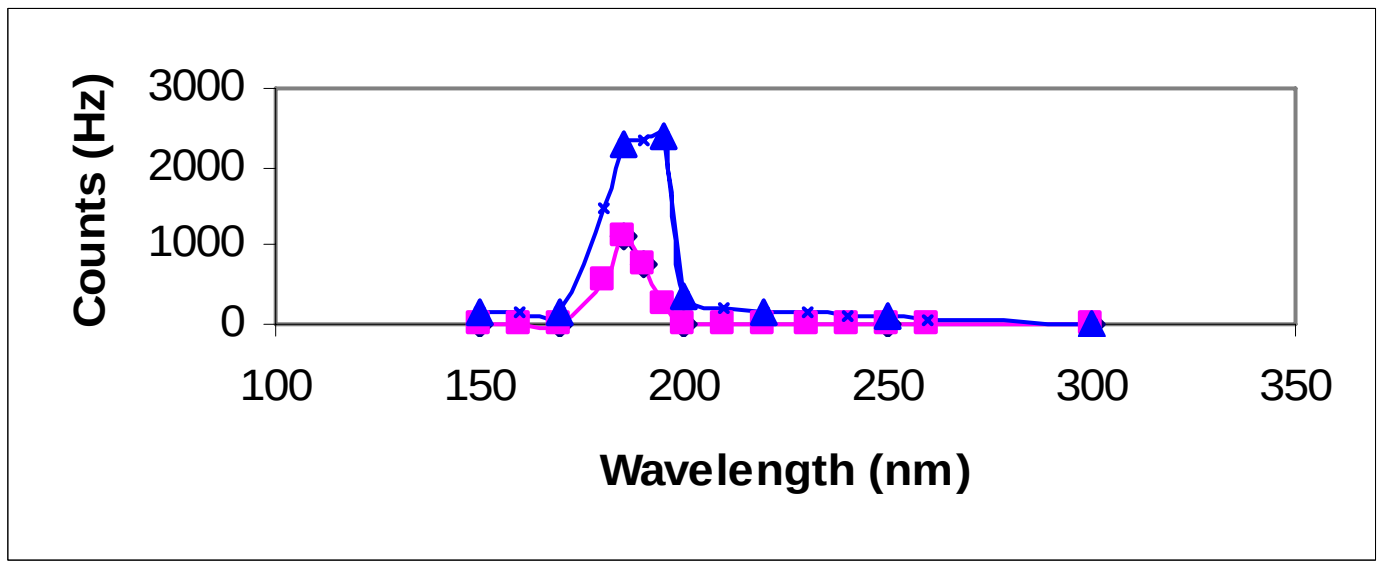

Fig.3. Spectra of the Hg lamp measured with the single-wire counter at $V_{TMAE}$=1940 V (triangles) and with double RETGEM at $V_{RETGEM}$=675 V(squares)

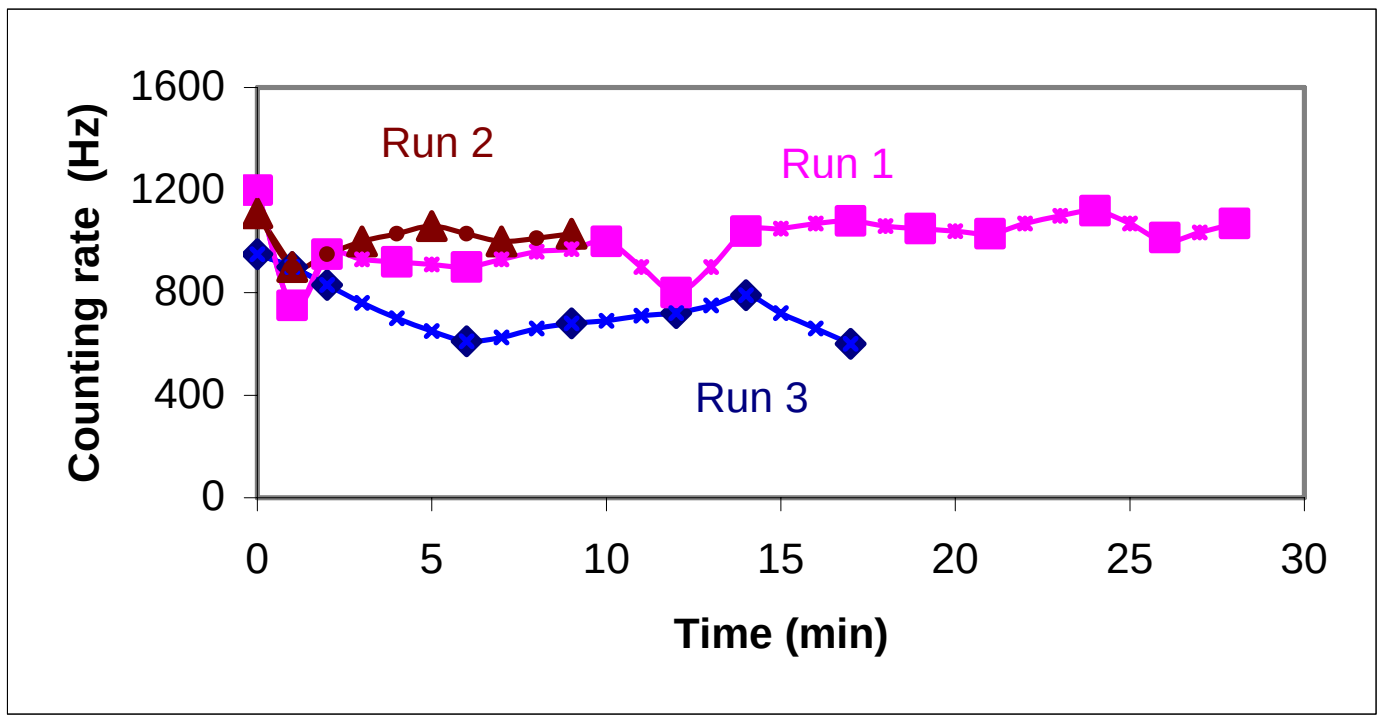

Fig. 4. Counting rate vs. time from the RETGEM irradiated by the UV light (185nm) from the monochromator. The light is concentrated on a small area of $6mm^2$.
About 30min without light have passed between each run, the scaler threshold was the same in each measurements

In the case of the flood illumination, when the full detector sensitive area ($5x5cm^2$) of the detector was illuminated by the UV light, the rate characteristics were considerably worse, as shown in Fig. 5. The similar effect was observed earlier for RPC detectors [7].
We have also performed a "long-term" stability test; the results obtained with two different CsI photocathodes are presented in Fig. 6. First one can see that two different photocathodes evaporated on the top electrode of RETGEM at different time gave identical QE. Second, no degradation with time was for both photocathodes indicated that they will probably remain stable and for a much longer period of time.

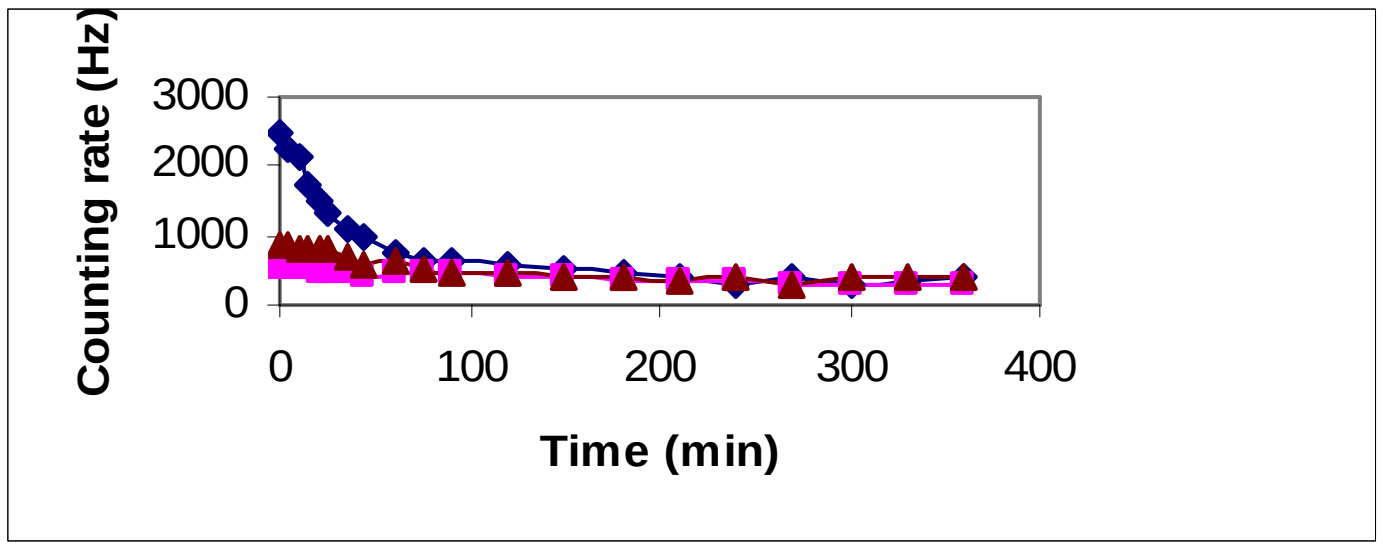

Fig. 5. Typical results obtained in the case of the flood illumination; one can note that the rate characteristics are worse compared to the case when the beam light was focused on a small area (see Fig. 4)

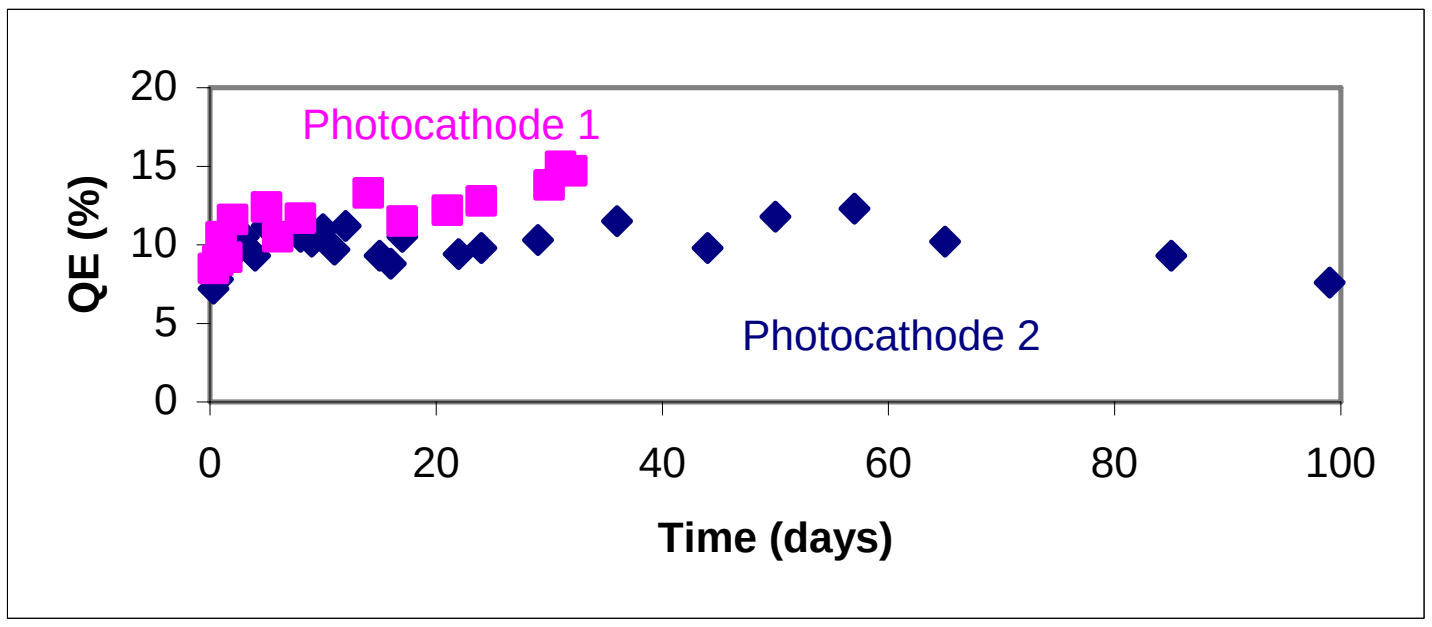

Fig.6. Long-term stability of the double RETGEM measured with two different CsI photocathodes evaporated on its top electrode

## Discussion and outlook

Preliminary results obtained in this work indicate that photosensitive RERGEM could be a candidate for VHMPID: it shows high enough QE and promising long term stability. The final evaluation of this detector will be done after designing and performing extensive tests with beam particles of a VHMPID prototype. The work in this direction has already started in our group.

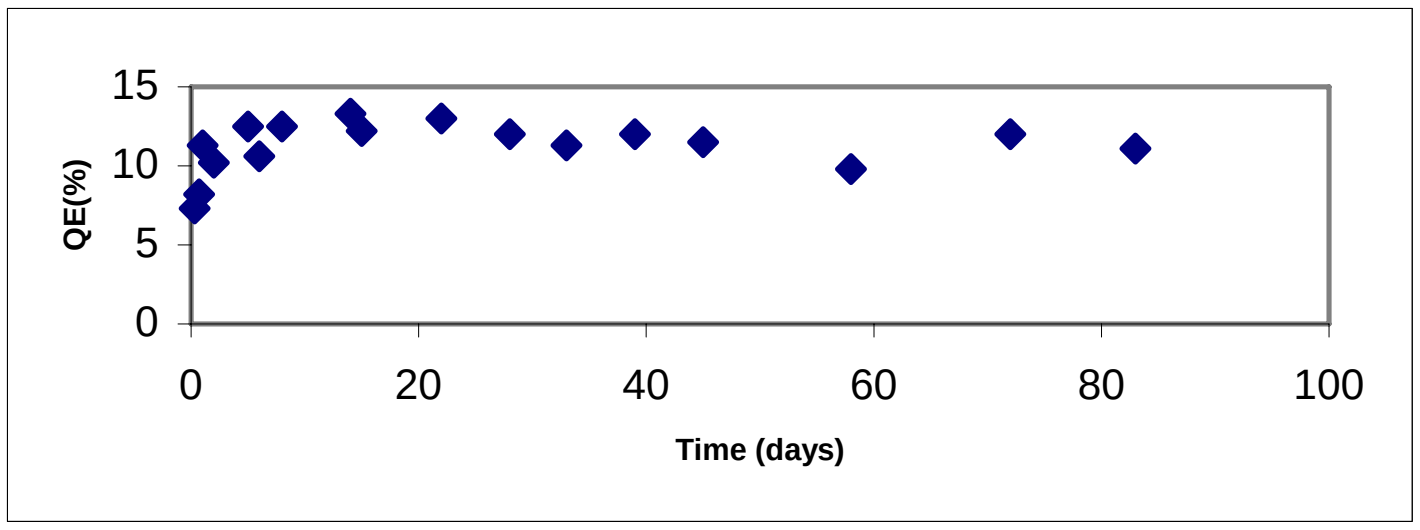

Fig. 7. Long-term stability test of a RETGEM manufactured by a screen-printing technology

Unfortunately, Dupon Company, which is the only producer of the resistive Kapton imposed some restriction for the non-US residents and it is not easy now to obtain this material. For this reason we have recently developed and tested RETGEMs manufactured by means of a screen-printing technology [8]. Screen printing technique is widely available in many labs and companies and allows production of large -area RETGEMs (up to ($50x50cm^2$) with a possibility to adjust the electrode resistivity to the requirements of the specific experiment. For example, it looks possible to considerably improve the rate characteristics of the RETGEM without losing its spark-protection property as it was already earlier done with RPC [9]. We already tested screen-printing RETGEM coated with a CsI layer. As one can see from Fig. 7 similar QE and stability with time were obtained with this new detector, which could be another possible candidate for the forthcoming VHMID detector.

## Acknowledgements

We would like to thank A. Braem, R. Oliveira, J. Van Beelien and M. Van Stenis for their help throughout this work. One of us (A. G. Agócs) was supported by grant NK62044 of the Hungarian OTKA .